\documentclass[twocolumn,showpacs,preprintnumbers,amsmath,amssymb,superscriptaddress,PRL]{revtex4}
\usepackage{graphicx}
\usepackage{dcolumn}
\usepackage{bm}
\usepackage{epsf} 

\usepackage{amsmath,amssymb}
\usepackage{graphicx}
\usepackage{amssymb}
\usepackage{epstopdf}
\usepackage{makeidx}
\begin{document}
\topmargin -1.5cm

\title{Search for the photo-excitation of exotic mesons in the $\pi^{+}\pi^{+}\pi^{-}$ system}

\newcommand*{\JLAB}{Thomas Jefferson National Accelerator Facility, Newport News, Virginia 23606}
\affiliation{\JLAB}
\newcommand*{\NSU}{Norfolk State University, Norfolk, Virginia 23504}
\affiliation{\NSU}
\newcommand*{\RPI}{Rensselaer Polytechnic Institute, Troy, New York 12180-3590}
\affiliation{\RPI}
\newcommand*{\FSU}{Florida State University, Tallahassee, Florida 32306}
\affiliation{\FSU}
\newcommand*{\ANL}{Argonne National Laboratory, Argonne, Illinois 60439}
\affiliation{\ANL}
\newcommand*{\ASU}{Arizona State University, Tempe, Arizona 85287-1504}
\affiliation{\ASU}
\newcommand*{\UCLA}{University of California at Los Angeles, Los Angeles, California  90095-1547}
\affiliation{\UCLA}
\newcommand*{\CSU}{California State University, Dominguez Hills, Carson, CA 90747}
\affiliation{\CSU}
\newcommand*{\CMU}{Carnegie Mellon University, Pittsburgh, Pennsylvania 15213}
\affiliation{\CMU}
\newcommand*{\CUA}{Catholic University of America, Washington, D.C. 20064}
\affiliation{\CUA}
\newcommand*{\SACLAY}{CEA-Saclay, Service de Physique Nucl\'eaire, 91191 Gif-sur-Yvette, France}
\affiliation{\SACLAY}
\newcommand*{\CNU}{Christopher Newport University, Newport News, Virginia 23606}
\affiliation{\CNU}
\newcommand*{\UCONN}{University of Connecticut, Storrs, Connecticut 06269}
\affiliation{\UCONN}
\newcommand*{\ECOSSEE}{Edinburgh University, Edinburgh EH9 3JZ, United Kingdom}
\affiliation{\ECOSSEE}
\newcommand*{\FU}{Fairfield University, Fairfield CT 06824}
\affiliation{\FU}
\newcommand*{\FIU}{Florida International University, Miami, Florida 33199}
\affiliation{\FIU}
\newcommand*{\GWU}{The George Washington University, Washington, DC 20052}
\affiliation{\GWU}
\newcommand*{\ECOSSEG}{University of Glasgow, Glasgow G12 8QQ, United Kingdom}
\affiliation{\ECOSSEG}
\newcommand*{\ISU}{Idaho State University, Pocatello, Idaho 83209}
\affiliation{\ISU}
\newcommand*{\INFNFR}{INFN, Laboratori Nazionali di Frascati, 00044 Frascati, Italy}
\affiliation{\INFNFR}
\newcommand*{\INFNGE}{INFN, Sezione di Genova, 16146 Genova, Italy}
\affiliation{\INFNGE}
\newcommand*{\ORSAY}{Institut de Physique Nucleaire ORSAY, Orsay, France}
\affiliation{\ORSAY}
\newcommand*{\ITEP}{Institute of Theoretical and Experimental Physics, Moscow, 117259, Russia}
\affiliation{\ITEP}
\newcommand*{\JMU}{James Madison University, Harrisonburg, Virginia 22807}
\affiliation{\JMU}
\newcommand*{\KYUNGPOOK}{Kyungpook National University, Daegu 702-701, South Korea}
\affiliation{\KYUNGPOOK}
\newcommand*{\MIT}{Massachusetts Institute of Technology, Cambridge, Massachusetts  02139-4307}
\affiliation{\MIT}
\newcommand*{\UMASS}{University of Massachusetts, Amherst, Massachusetts  01003}
\affiliation{\UMASS}
\newcommand*{\MOSCOW}{Moscow State University, General Nuclear Physics Institute, 119899 Moscow, Russia}
\affiliation{\MOSCOW}
\newcommand*{\UNH}{University of New Hampshire, Durham, New Hampshire 03824-3568}
\affiliation{\UNH}
\newcommand*{\OHIOU}{Ohio University, Athens, Ohio  45701}
\affiliation{\OHIOU}
\newcommand*{\ODU}{Old Dominion University, Norfolk, Virginia 23529}
\affiliation{\ODU}
\newcommand*{\PITT}{University of Pittsburgh, Pittsburgh, Pennsylvania 15260}
\affiliation{\PITT}
\newcommand*{\RICE}{Rice University, Houston, Texas 77005-1892}
\affiliation{\RICE}
\newcommand*{\URICH}{University of Richmond, Richmond, Virginia 23173}
\affiliation{\URICH}
\newcommand*{\SCAROLINA}{University of South Carolina, Columbia, South Carolina 29208}
\affiliation{\SCAROLINA}
\newcommand*{\UTFSM}{Universidad T\'ecnica Federico Santa Mar\'{\i}a, Casilla 110-V, Valpara\'\i so, Chile}
\affiliation{\UTFSM}
\newcommand*{\UNIONC}{Union College, Schenectady, NY 12308}
\affiliation{\UNIONC}
\newcommand*{\VT}{Virginia Polytechnic Institute and State University, Blacksburg, Virginia   24061-0435}
\affiliation{\VT}
\newcommand*{\VIRGINIA}{University of Virginia, Charlottesville, Virginia 22901}
\affiliation{\VIRGINIA}
\newcommand*{\WM}{College of William and Mary, Williamsburg, Virginia 23187-8795}
\affiliation{\WM}
\newcommand*{\YEREVAN}{Yerevan Physics Institute, 375036 Yerevan, Armenia}
\affiliation{\YEREVAN}
\newcommand*{\NOWLANL}{Los Alamos National Laboratory, Los Alamos, New Mexico 87545}
\newcommand*{\NOWOHIOU}{Ohio University, Athens, Ohio  45701}
\newcommand*{\NOWJLAB}{Thomas Jefferson National Accelerator Facility, Newport News, Virginia 23606}
\newcommand*{\DECEASED}{.}
\newcommand*{\NOWUNH}{University of New Hampshire, Durham, New Hampshire 03824-3568}
\newcommand*{\NOWCNU}{Christopher Newport University, Newport News, Virginia 23606}
\newcommand*{\NOWSCAROLINA}{University of South Carolina, Columbia, South Carolina 29208}
\newcommand*{\NOWUMASS}{University of Massachusetts, Amherst, Massachusetts  01003}
\newcommand*{\NOWMIT}{Massachusetts Institute of Technology, Cambridge, Massachusetts  02139-4307}
\newcommand*{\NOWURICH}{University of Richmond, Richmond, Virginia 23173}
\newcommand*{\NOWECOSSEE}{Edinburgh University, Edinburgh EH9 3JZ, United Kingdom}
\newcommand*{\NOWGEISSEN}{Physikalisches Institut der Universitaet Giessen, 35392 Giessen, Germany}

\author {M.~Nozar} 
\affiliation{\JLAB}
\author {C.~Salgado} 
\affiliation{\NSU}
\author {D.P.~Weygand} 
\affiliation{\JLAB}
\author {L.~Guo} 
\altaffiliation[Current address:]{\NOWLANL}
\affiliation{\JLAB}
\author {G.~Adams} 
\affiliation{\RPI}
\author {Ji~Li} 
\affiliation{\RPI}
\author {P.~Eugenio} 
\affiliation{\FSU}
\author {M.J.~Amaryan} 
\affiliation{\ODU}
\author {M.~Anghinolfi} 
\affiliation{\INFNGE}
\author {G.~Asryan} 
\affiliation{\YEREVAN}
% ...
\collaboration{The CLAS Collaboration}
     \noaffiliation

\date{\today}

\begin{abstract}
A search for exotic mesons in the  $\pi^{+}\pi^{+}\pi^{-}$ system photoproduced by the charge exchange reaction 
$\gamma p\rightarrow \pi^{+}\pi^{+}\pi^{-}(n)$ was carried out by the CLAS collaboration at Jefferson Lab.  A tagged-photon beam with energies in the 4.8 to 5.4 GeV range, produced
through bremsstrahlung from a 5.744 GeV electron beam, was incident on a liquid-hydrogen target.  A Partial Wave Analysis (PWA)  was performed on a sample of 83,000 events, the highest such statistics to date in this reaction at these energies.  The main objective of this study was to look for the photoproduction of an exotic $J^{PC} = 1^{-+}$ resonant state in the 1 to 2 GeV mass range. Our PWA analysis, based on the isobar model, shows production of the $a_{2}(1320)$
and the $\pi_{2}(1670)$ mesons, but no evidence for the $a_{1}(1260)$, nor the $\pi_{1}(1600)$ exotic
state at the expected levels.  An upper limit of 13.5 nb is determined for the 
exotic $\pi_1(1600)$ cross section, less than $2\%$ of the $a_2(1320)$ production.
\end{abstract}

\pacs{13.60.Rj, 13.60.-r, 14.60.jn}

\maketitle

The self interacting nature of the gluon within Quantum Chromodynamics (QCD) allows for hybrid resonant states with a $(\bar{q}qg)$ configuration, where the gluonic degree of freedom  gives rise to a spectrum of additional states outside the constituent quark model (CQM). The observation of gluonic-hybrid hadrons, with an explicit excitation of this gluonic degree of freedom, will be an important test of the predicting power of QCD at intermediate energies. A signature of states beyond the CQM would be the existence of mesons with quantum numbers that cannot be attained by ($\bar{q}q$) mesons (so-called ``exotics'').  In particular, the lowest lying $(\bar{q}qg)$ state is predicted to have  $J^{PC} = 1^{-+}$~\cite{BERNARD97,Juge,Lacock,Mei}, and mass near $1.9$ GeV. A more recent calculation on the lattice using lighter quark masses predicts a mass at $1.74$ GeV~\cite{hedditch}.

Here we report on a search for exotic mesons decaying to three charged pions. This channel was chosen for its simplicity, since only a few decay channels are available to this final state. In addition, although the dominant decay mode of the lightest $J^{PC}=1^{-+}$ state is predicted to be into an S-wave and P-wave meson, such as the $b_{1}(1235) \: \pi$ or $f_{1}(1285)\:\pi$~\cite{nathan}, the three-pion final state in the $\rho$ decay channel could be non-negligible~\cite{close1,close2,page}.

There is also evidence for an exotic $\pi_1(1600)$ state in the reaction $\pi^- p \rightarrow \pi^{+}\pi^{+}\pi^{-}p$ at 18 GeV by the Brookhaven E852 experiment~\cite{Adams, chung} . Furthermore, the existence of the $\pi_1(1600)$ has been confirmed by the same experiment in other channels such as $\eta'\pi$~\cite{ivannov}. The VES collaboration also reported an exotic signal in the $\eta'\pi^-$ channel in the same mass region~\cite{VES}. However, a more recent analysis of a 
higher statistics sample from E852 3$\pi$ data, claims to find no evidence for  $\pi_1(1600)$ production~\cite{alex}. All these results were obtained by pion beam experiments. 

It has long been anticipated that photoproduction may be a better
production mechanism for exotic
mesons~\cite{nathan,close2,andre,andreadam,adam}.  If the $\pi_1(1600)$
state couples to $\rho \pi$, then this state should also be produced
with a photon beam through $\pi$ exchange via Vector-Meson-Dominance
(VMD)~\cite{gellmann}. However, the use of photon beams as probes for
exotic meson production has not been fully explored so far, and the
existing data on multi-particle final states are
sparse. Photoproduction of mesons in multi-pion final states has been
reported in three previous experiments: a SLAC 40-in. hydrogen bubble
chamber experiment~\cite{eisen2} at 4.3 GeV,  a CERN hydrogen
experiment~\cite{aston} between 25 and 70 GeV, and a SLAC 1-meter
hybrid bubble chamber experiment~\cite{condo2,condo1} around 19 GeV.
All these experiments lacked the statistics required for a full
Partial Wave Analysis.  For example, Condo {\it et al.}~\cite{condo2} showed that the three-pion spectrum in the low-mass region (below 1.5 GeV) is dominated by $a_{2}(1320)$ production, and found no clear evidence for $a_{1}(1260)$ production. In the high-mass region (1.5 to 2 GeV), based on an angular distribution analysis, this same group claimed evidence for a narrow state at 1.775 GeV with possible $J^{PC} = 1^{-+}$, $2^{-+}$, or $3^{++}$ assignments. From the analysis of four pion events, Ref.~\cite{aston} reported two peaks, one at the mass of the $a_{2}(1320)$ and another at around 1.75 GeV, in the $\rho\pi$  state recoiling off the proton and the remaining pion.

The present experiment was performed at Jefferson Laboratory during the 2001 running period of CLAS. Details on the design and
 operation of CLAS and its components may be found in 
 Ref.~\cite{bernard} and references within. The experiment ran with a fixed electron beam energy of 5.744 GeV. A
tagged-photon beam, with a flux of $5 \times 10^7$ photons/s produced
via bremsstrahlung from a  3x$10^{-4}$ radiation lengths radiator, was
incident on a liquid-hydrogen target contained in a cylindrical cell
18 cm in length. The running conditions were optimized for meson
production recoiling off a neutron: the CLAS torus magnet field was
reduced to half its maximum current, and the target pulled back 1~m
from the nominal center of CLAS position. Only events with
tracks coming from the target material that were in time within 1 ns
with a beam photon with energy between 4.8 and 5.4 GeV were selected.

\begin{figure}[htbp!]
\includegraphics[width=3.3in,height=2.4in]{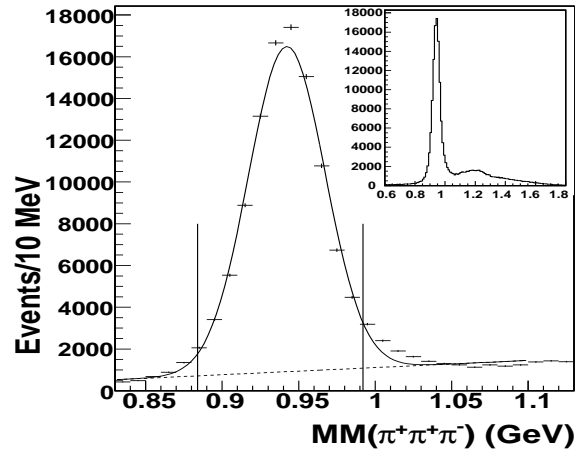}
\caption{Missing mass of the three-pion system. The distribution was fit to a gaussian plus a linear background (solid curve). The inset shows a wider scale. The region between the vertical lines at $0.884$ and $0.992$ GeV identified the neutron.}
\label{fig:mm}
\end{figure}

\begin{figure}[htbp!]
\includegraphics[width=3.4in]{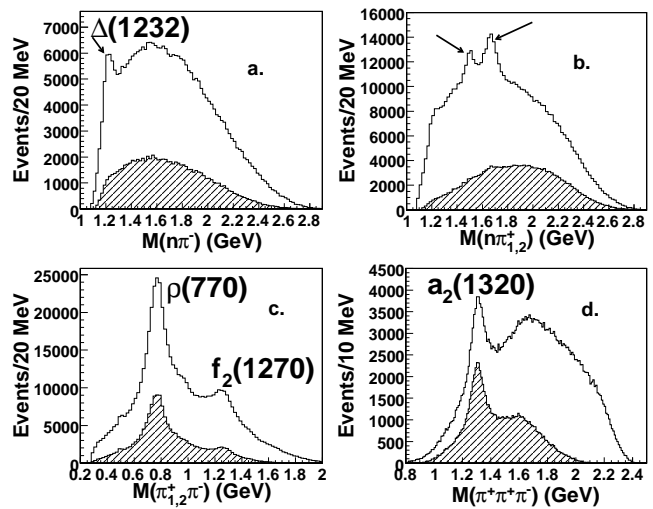}
\vspace{-5mm}
\caption{Mass spectra. (a) $n \pi^-$ invariant mass, (b) $n \pi^+$ invariant mass (the left arrow indicates N*(1520/1535) resonances and the right arrow  N*(1650/1675/1680) resonances), (c) $\pi^+\pi^-$ invariant mass, and (d) $\pi^+ \pi^+ \pi^-$ invariant mass. The hatched areas are the distributions
of remaining events after the momentum  transfer and $\pi^+$ laboratory angle selections. The histograms in (b) and (c) are filled twice for each event, once for each $\pi^+$.}
\label{fig:masses}
\end{figure}

\begin{figure}[htbp!]
\includegraphics[width=3.0in]{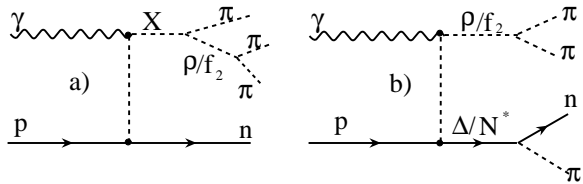}
\caption{(a) Peripheral meson production, (b) baryon resonance background processes.}
\label{fig:diag}
\end{figure}

The three pions  in the reaction $\gamma p\rightarrow \pi^+ \pi^+ \pi^- n$ were identified by the time-of-flight detector, and the neutron was identified through missing mass (Fig.~\ref{fig:mm}). The reconstructed mass of the neutron was $942.2$ MeV with a width of $25.1$ MeV. As shown in Fig.~\ref{fig:masses}$a, b$, the various $n\pi$ invariant mass spectra show clear production of baryon resonances recoiling off a two-pion system, consistent with the process shown in Fig.~\ref{fig:diag}$b$. The $n\pi^{-}$ invariant mass  (Fig.~\ref{fig:masses}$a$) shows a peak at the
mass of the $\Delta(1232)$, while the $n\pi^{+}$  effective mass distribution (Fig.~\ref{fig:masses}$b$) shows enhancements around the mass of N*(1520/1535) and N*(1650/1675/1680), with
a smaller contribution from $\Delta(1232)$ production.

The observed squared four-momentum transfer distribution, $-t$, between the incoming photon
beam and the produced three-pion system follows an exponential of the form $f(t) = a e^{-b|t|}$.  This form is consistent with peripheral production of the three-pion system recoiling off the
neutron and is consistent with the production mechanism shown in Fig.~\ref{fig:diag}$a$. For the exponential slope constant, $b$, we obtained a fitted value of $3.02$ GeV$^{-2}$. To enrich the sample of peripheral events, we required $|t| \leq 0.4$  GeV$^{2}$. To further enhance the mesonic sample relative to the baryonic background, only events with forward-going $\pi^{+}$'s ($\theta_{Lab}(\pi^{+}) \leq 30^{o}$) were selected. This requirement, as determined from simulation, is necessary to remove most of the baryonic background without significantly affecting the mesonic sample. After all the former selections were applied, about 83,000 events remained as input to the Partial Wave Analysis (PWA).

The $\pi^{+}\pi^{-}$ invariant mass distribution, Fig.~\ref{fig:masses}$c$, shows enhancements at the mass of the $\rho (770)$ and $f_{2}(1270)$. The 3$\pi$ invariant mass spectrum, Fig.~\ref{fig:masses}$d$, shows a clear peak around the mass of the $a_{2}(1320)$ meson. It also shows a broad enhancement in the 1600 - 1700 MeV mass region, which could be due to the $\pi_2(1670)$ or $\pi_1(1600)$, but it could also be the result of acceptance and phase space limitations.

A Partial Wave Analysis, based on the isobar model, was performed to
determine the spin, parity, and charge-conjugation ($J^{PC}$) of the three-pion system. The three-body decay of the meson was described by a
sequence of two-body intermediate decays
(Fig.~\ref{fig:diag}$a$). States with definite $J^{PC}$ of the mesonic
system were combined as eigenstates of reflection through the
production plane (reflectivity) according to the formalism described
in Refs.~\cite{chung2,trueman,weygand,eugenio}.  The expected
number of events per mass bin were written as a sum of production and
decay amplitudes. To obtain the observed number of events per bin, the
acceptance of the detector was determined by normalizing to the Monte
Carlo sample. The model used the same $t$ distribution as determined
from the data, and the events were simulated in a GEANT-based model of
the CLAS detector. A maximum likelihood fit to each mass bin was done with
a set of input partial waves. Many different sets of waves were tried during the analysis. The meson spin-density matrix was approximated by a rank 2 matrix due to limited statistics and the large number of parameters required for the fit. The rank of the spin-density matrix represents the number of independent spin components in the initial and final states necessary to describe the interaction. Parity conservation reduces the number of independent components to four. 

No baryon resonance partial waves were used and an isotropic non-interfering background wave was included, which is appropriate, since the events were dominated by meson production after the kinematic selection cuts. Below $1.36$~GeV, no $f_2\pi$ wave was included as its mass is lower than the nominal threshold. Contributions from  $m^{\epsilon} = 0^{+},0^{-}$ waves were found to be small, and these waves are therefore not included in the final wave set. The final list of waves that provided the best stable fit is shown in Table~\ref{tab:ListOfWaves}. The quality of the PWA fit was checked by comparing various mass and angular distributions
from the observed data and the predictions of the PWA results, i.e., weighting the Monte Carlo events according to the PWA fit. Qualitatively good agreement was observed between these two sets~\cite{mina}.

\begin{table}[ht!]
\vspace*{6pt}
\begin{tabular}{|c|c|l|l|c|c|}
\hline
$\pi^{+}\pi^{+}\pi^{-}$ mass & $J^{PC}$             & $m^{\epsilon}$                & {L}       & { Isobar}     & {$\#$ Waves}\\\hline\hline
$1.0-1.36$ GeV                              & $1^{++}$             & {$1^{\pm}$}                   & {$0$,$2$} & $\rho(770)$   & $4$  \\ \cline{2-6}
       & $2^{++}$                 & {$1^{\pm}$,$2^{\pm}$}         & $2$       & $\rho(770)$   & $4$  \\ \cline{2-6}
  & $1^{-+}$             & {$1^{\pm}$}                   & $1$       & $\rho(770)$   & $2$  \\ \cline{2-6}
                            & $2^{-+}$             & {$1^{\pm}$}                   & $1$       & $\rho(770)$   & $2$  \\ \cline{2-6}
                                            & { Flat bg}           &\multicolumn{3}{c}{}           &       \multicolumn{1}{|c|}{$1$}  \\ \hline
$1.36-2.0$ GeV                              & $1^{++}$             & {$1^{\pm}$}                   & {$0$,$2$} & $\rho(770)$   & $4$  \\ \cline{2-6}
                                  & $1^{++}$             & {$1^{\pm}$}                   & $1$       & $f_{2}(1270)$ & $2$  \\ \cline{2-6}
  & $2^{++}$             & {$1^{\pm}$,$2^{\pm}$}         & $2$       & $\rho(770)$   & $4$  \\ \cline{2-6}
                           & $1^{-+}$             & {$1^{\pm}$}                   & $1$       & $\rho(770)$   & $2$  \\ \cline{2-6}
                                            & $2^{-+}$             & {$1^{\pm}$,$2^{\pm}$}         & $1$       & $\rho(770)$   & $4$  \\ \cline{2-6}
                                            & $2^{-+}$             & {$1^{\pm}$}                   & $3$       & $\rho(770)$   & $2$  \\ \cline{2-6}
                                            & $2^{-+}$             & {$1^{\pm}$,$2^{\pm}$}         & $0$       & $f_{2}(1270)$ & $4$  \\ \cline{2-6}
                                            & $2^{-+}$             & {$1^{\pm}$}                   & $2$       & $f_{2}(1270)$ & $2$  \\ \cline{2-6}
                                            & $3^{++}$             & {$1^{\pm}$}                   & $2$       & $\rho(770)$   & $2$  \\ \cline{2-6}
                                            & $3^{++}$             & {$1^{\pm}$}                   & $1$       & $f_{2}(1270)$   & $2$  \\ \cline{2-6}
                                            & { Flat bg}           &\multicolumn{3}{c}{}           &        \multicolumn{1}{|c|}{$1$} \\\hline
\end{tabular}
\caption{Set of partial waves used in the PWA. $J, P, C$ refers to the Spin, Parity, Charge-conjugation of the three-pion system. $m, \epsilon$ are the spin-projection and reflectivity quantum numbers. L is the orbital angular momentum of the decay. Isobar is the intermediate meson that decays to two pions.}
\label{tab:ListOfWaves}
\end{table}

\begin{figure}[htbp!]
\includegraphics[width=3.4in]{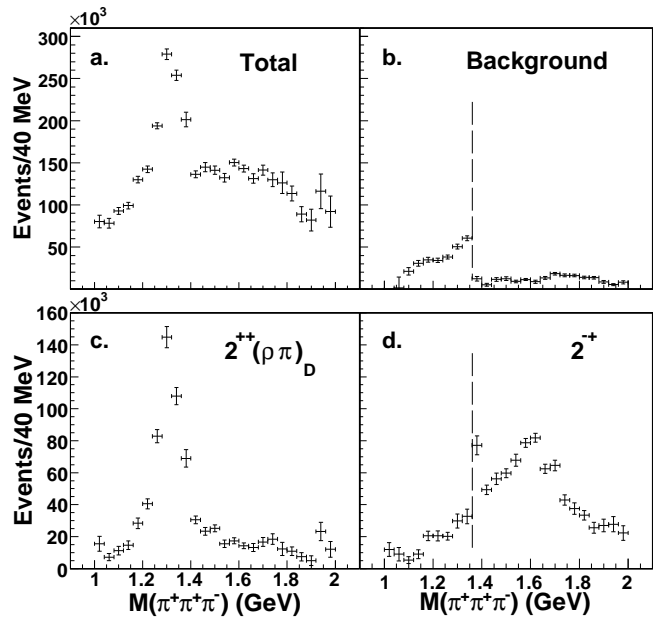}
\caption{PWA results: Combined intensities of waves included in the fit.  The intensities shown are the sum of the intensities from both ranks of the spin-density matrix: (a) Total intensity, (b) Background intensity, (c) $2^{++}(\rho \pi)_D$ intensity, and (d) $2^{-+}$ total intensity. The dashed vertical line in (c) and (d)) indicates the border of the two fitting regions.}
\label{fig:pwa1}
\end{figure}

\begin{figure}[htbp!]
\includegraphics[width=3.4in]{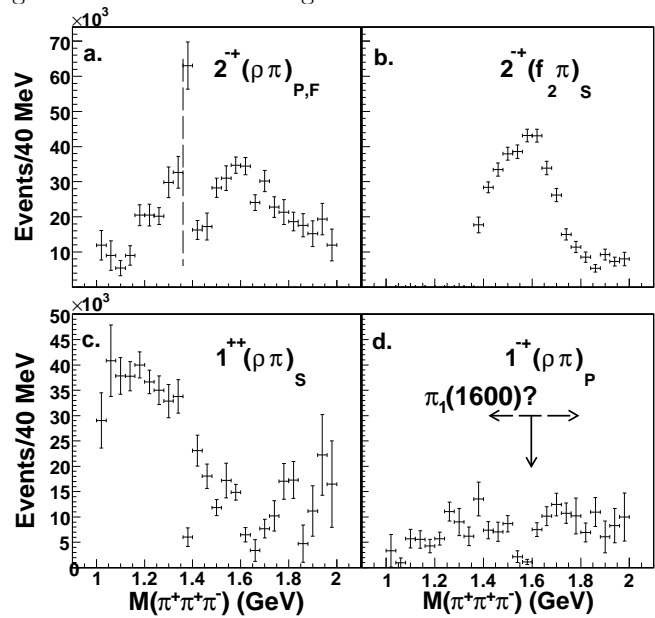}
\caption{Partial wave decompositions: (a) $2^{-+}(\rho\pi)_{P,F}$ (the dashed vertical line indicates the border of the two fitting regions), (b) $2^{-+}(f_2\pi)_{S}$, (c) $1^{++}(\rho\pi)_{S}$, and (d) $1^{-+}(\rho\pi)_{P}$ ( the vertical and horizontal arrows indicate the mass and width of the $\pi_1(1600)$ as reported by Ref.~\cite{Adams}). }
\label{fig:pwa2}
\end{figure}

Results from the PWA are shown in Figs.~\ref{fig:pwa1}, and \ref{fig:pwa2}. The two most prominent waves are  $J^{PC}=2^{++}$ and $2^{-+}$ (Fig.~\ref{fig:pwa1}$c, d$). The $a_{2}(1320)$ is observed in the $(\rho \pi)_D$ wave at the expected mass and width. 
For the $a_{2}$ cross section, we obtained a value of $0.81 \pm 0.25 \: \mu b$. The uncertainty includes both detector systematics and variation of PWA results estimated by using different wave sets and starting parameters. The result is in rough agreement with
the previous measurement of $1.14 \pm 0.57 \: \mu b$ obtained at 5.25 GeV by Ballam~\cite{ballam} and $1.71 \pm 0.86 \: \mu b$ obtained at
4.7 to 5.8 GeV by Eisenberg~\cite{eisen}.

The $\pi_{2}(1670)$ is observed clearly in the $(f_{2}\pi)_{S}$ decay
mode (Fig.~\ref{fig:pwa2}$b$). In the remaining $\pi_{2}$ decay modes: $(\rho\pi)_{P,F}$(Fig.~\ref{fig:pwa2}$a$), and $(f_{2}\pi)_{D}$, the signal looks
broad and distorted in the low-mass region. This is most likely due to the incomplete set of waves in the fit, and is indicative of leakage from the $a_2$. Similar observations were reported by the E852 collaboration~\cite{chung,alex}. In addition, due to the two different wave sets used in the two fitting regions (Table~\ref{tab:ListOfWaves}), some instability is observed near the border of 1.36~GeV (Fig.~\ref{fig:pwa1}$b$, $d$). This should be expected as a result of incomplete waves sets and can only be resolved by future higher statistics experiments~\cite{superg}.

It is important to note that we see no clear evidence for the $a_{1}(1260)$ in the possible decay mode of the $(\rho\pi)_{S}$ wave (Fig.~\ref{fig:pwa2}$c$). This observation agrees with Condo et al.~\cite{condo1}. The event enhancement observed around 1.3 GeV in the non-interfering,
flat background waves (Fig.~\ref{fig:pwa1}$b$) is likely leakage
coming from the $a_{2}$ and may demonstrate that the rank of the spin density matrix is larger than 2.
 
We do not observe resonant structure in the exotic $1^{-+} (\rho
\pi)_P$ partial wave (Fig.~\ref{fig:pwa2}$d$). 
   To determine if the $1^{-+}$ wave was necessary to better fit our
   data, we compared PWA fits with and without the $1^{-+}$ wave using
   the Likelihood Ratio (LR) test~\cite{LR1,LR2}. Using LR statistics, we
   find that the PWA set of waves including the $1^{-+}$ wave fits the
   data significantly better than a model without the 
   $1^{-+}$ wave.  
While no clear resonant structure was
   observed in the $1^{-+}$ intensity distribution
   (Fig.~\ref{fig:pwa2}$d$), this distribution was used to estimate an
   upper limit to the $\pi_1(1600)$ cross section using the method 
   of ~O.~Helene~\cite{helene}.
Using the mass of $1597$~MeV and the width of $340$~MeV as measured by Ref.~\cite{ivannov}, we estimated an upper limit for the $\pi_1(1600)$ of
13.5 nb at a 95\% confidence level, less than 2\% of the
$a_2(1320)$. Therefore, our results do not agree with the predicted 
strengths for photoproduction of a $1^{-+}$ gluonic hybrid
meson~\cite{nathan,close2,andre,andre2,adam}. Based on Ref.~\cite{close2},
the $\pi_1(1600)$ is expected to be produced with a strength near $10\%$ of the
$a_2(1320)$. Ref.~\cite{adam} predicts a factor of 5-10 larger ratio
of exotic meson to $a_2$ in photoproduction than hadroproduction. This
would imply the $\pi_1(1600)$ cross section to be on the order of
50\% of $a_2$, more than 25 times higher than what we observed. It is
possible that the $\pi_1(1600)$ reported by Ref.~\cite{Adams} is not a
gluonic hybrid meson, but rather of other nature. Alternatively, the
calculated photoproduction cross section of gluonic exotic mesons
might be overestimated.

This work was supported in part by the U.S. Department of Energy,
the U.S. National Science Foundation, the Italian Istituto Nazionale di Fisica Nucleare,
the  French Centre National de la Recherche Scientifique,
the French Commissariat \`{a} l'Energie Atomique, and the Korean Science and Engineering Foundation.
Jefferson Science Associates (JSA) operates the
Thomas Jefferson National Accelerator Facility for the United States
Department of Energy under contract DE-AC05-060R23177.

\end{document}